\documentstyle[ichep,12pt]{article}
\def\ie{{\it i.e.}}
\def\eg{{\it e.g.}}
\newif\ifproceeding
\proceedingtrue
\title { Stability of
 Vacua and Domain Walls in Supergravity and Superstring
Theory\ifproceeding\thanks{ Invited Talk Presented at the XXVI
International Conference on High Energy Physics, August 6-12, 1992,
Dallas, Texas.}\fi} \author {Mirjam Cveti\v
c\ifproceeding\thanks{email: cvetic@cvetic.hep.upenn.edu}\fi\\ Physics
Department \\ University of Pennsylvania, Philadelphia PA 19104-6396}
\begin{document}
\finalcopy
\ifproceeding\pagestyle{myheadings}
\markright{\vbox{\rm\noindent UPR--529--T\hfill\vskip 3mm\noindent September,
1992\hfill\vskip 4mm}}\fi
\maketitle
\abstract
{We address the possibility of false vacuum decay in $N=1$
supergravity theories, including those corresponding to superstring
vacua. By establishing a
Bogomol'nyi bound for the energy density stored in the
domain wall of
the $O(4)$ invariant bubble, we show that supersymmetric vacua
remain absolutely stable against false vacuum decay
into another supersymmetric vacuum, including those from a
Minkowski to an anti-deSitter (AdS)
one. As a consequence,
 there are no compact
static spherical domain walls, while on the other hand there
exist planar domain walls interpolating between
non-degenerate supersymmetric vacua, \eg   between
Minkowski (topology $\Re^{4}$)
and AdS (topology $S^{1}(time) \times \Re^{3}(space)$) vacua.}

\vskip -1pc

The mechanism of false vacuum decay is presently
well understood\cite{COL}.
Quantum-mechanical tunnelling from a false vacuum
to a true one proceeds via formation of a true vacuum
bubble inside the false vacuum
background. Energy gained by forming the true vacuum bubble
less the wall energy is transferred to the wall kinetic energy,
driving the wall expand asymptotically to the speed of light.
The effects of gravity on the false vacuum decay we studied by
Coleman and DeLuccia\cite{COLDL}
In particular, they found that a Minkowski
false vacuum cannot decay into an anti-de Sitter (AdS)
true vacuum unless the matter vacuum energy difference
$\Delta V$ is sufficiently large.  The result can be
generalized to
vacuum decay between two AdS vacua with
corresponding matter vacuum energies
$ V_{true}<V_{false}<0$.\vfil\eject
 In either case
\begin{equation}
 (\sqrt{-V_{true}}-\sqrt{-V_{false}})^2\ge
{3 \over 4}\kappa \sigma^2
\label{cdlboundII}
\end{equation}
must be satisfied in order for  a true vacuum
bubble to form.
Here, $\kappa\equiv 8\pi G_N$ and
 $\sigma $ denotes the bubble wall energy per unit area.

The study of false vacuum decay in $N=1$
supergravity theories is interesting on its own.
However, when the matter fields
are associated with the low energy   scale, the
splitting of the non-degenerate supersymmetric vacua
is  generically small; \ie , $\sim \kappa$.
 The effective Lagrangian
 of superstring vacua
 is described by an $N=1$ supergravity as well.
However, in superstring
theories, other fields, \eg , dilaton and  moduli,
and gravity are on an equal footing,
so the effects of gravity can yield
distinctly new features. In particular, when the vacuum
expectation values (VEV's) of the moduli fields   are
of the order of $M_{pl}$, the  non-perturbatively
induced potential of moduli fields
is significantly modified by gravity.
In this case $\Delta V ={\cal O}(V)$ and prior to this
study\cite{CGRII}
not much could be said about the
stability of such superstring vacua.

We point out that  quantum tunnellings
between any supersymmetric vacua  in $N=1$ supergravity
are absolutely impossible\cite{CGRII}
by establishing a Bogomol'nyi bound for the bubble wall energy density.
 In particular, vacuum decay
from a supersymmetric Minkowski vacuum to an AdS
supersymmetric vacuum is not possible at all.
This  in turn applies to the case of tunnelling in the
moduli sector of string theory
when the non-perturbative moduli
potential is turned on.

We con\-sid\-er four-dimensional $N=1$ local supersymmetry with
one chiral superfield $T$.
The bosonic sector of the  four-dimensional
$N=1$ supergravity Lagrangian reads
\begin{equation}
e^{-1} L = -{1 \over 2 \kappa} R + K_{T \bar T} g^{\mu \nu}
\nabla_\mu T \nabla_\nu \bar T - V(T, \bar T)
\label{lagran}
\end{equation}
in which the supergravity scalar potential $V(T, \bar T)$
is defined as
\begin{equation}
V \equiv e^{\kappa K} [ K^{T \bar T} |D_T W|^2
- 3\kappa|W|^2]
\label{potential}
\end{equation}
here
$e = |detg_{\mu \nu}|^{1 \over 2},
K(T, \bar T) $ is K\"ahler potential and $D_{T}W \equiv
e^{-\kappa K} (\partial_T e^{\kappa K}W)$.
Newton's constant appears consistently in the combination
$\kappa = 8 \pi G_{N}$.
Supersymmetry preserving minima of the scalar potential
(\ref{potential}) satisfy
$D_{T}W=0$. This in turn implies (see eq. (\ref{potential}))
that the supersymmetry preserving vacua have
either zero vacuum energy (Minkowski space-time)
when $ W=0$, or constant negative vacuum energy
$- 3\kappa e^{\kappa K} |W|^2$  (AdS space-time) when
$W\not=0$.  Thus, the tunnelling process between
supersymmetric vacua corresponds to tunnelling either
between Minkowski and AdS space-times or
between two AdS space-times of different cosmological
constants.

In superstring theories,
the scalar field $T$ \eg ,
corresponds to a modulus field
arising from compactification, and its non-perturbatively induced
superpotential $W$ is assumed to reflect
the underlying target space modular invariance\cite{FLST,CFILQ}
 under the $PSL(2,{\bf Z})$
duality transformations:
\begin{equation}
T\rightarrow{{aT-ib}\over{icT+d}}  \ \ \ ,
 ad-bc=1\ \ ,
 \{a, b, c, d\}\in {\bf Z}.\label{modtrans}
\end{equation}
In this case\cite{CFILQ} ,
$
K = - 3 \kappa\ln [(T + \bar T)]$.
The superpotential $W$ is a modular  function  of weight
$-3$ under  $PSL(2, \bf Z)$ defined in the fundamental
domain $\cal D$ of the $T$-field.
One of the simplest choices for a
modular invariant superpotential is
 $W(T) = j(T)\, \eta^{-6}(T) $
where $\eta (T)$ is the Dedekind function: a modular
function of weight $1/2$ and
$j(T)$ is  a modular-invariant function.\cite{SCH}
In this case\cite{CFILQ}
 the  scalar
potential for the  $T$ field has two  supersymmetric
minima at $T=1$ ($V<0$)
and $T=\rho \equiv e^{i\pi/6}$ ($V=0$), both of which
lie on the boundary of the fundamental domain ${\cal D}$.
This is an example of two supersymmetric non-degenerate
superstring vacua.

The bubble formation  is studied
by using   an $O(4)$ symmetric Ansatz\cite{COL}
for a bounce solution interpolating between the
true(AdS) and false(Minkowski or AdS) vacuum.  The
metric for this Ansatz in Euclidean space
is\cite{COLDL}
\begin{eqnarray}
ds^{2}& =& d\xi^{2} + R(\xi)d\Omega_{3}^{2}\nonumber\\
&=&B(\xi')(d\tau^2+dr^2+r^2d\Omega_2^2)
\label{ofourm}
\end{eqnarray}
where $\xi$ is the Euclidean radial distance from
an arbitrary origin and $\xi '^2=\tau^2+r^2$.
The second line of (\ref{ofourm}) follows after a
redefinition of the radial coordinate $\xi$ into
$\xi'$: $d\xi '/\xi '=d\xi/\sqrt{R(\xi)}$.
Note that only with  coordinate
  $\xi '$   we can clearly attach the meaning to
$r=\sqrt{x^2+y^2+z^2}$ as the radius of the two sphere.
 The classical evolution of the materialized bubble
is described by the Wick rotation back to Minkowski
space-time,
\ie , by changing the Euclidean time $\tau$ back to
Minkowski time $t$.

It is most convenient to study the energy density of
the bubble wall at the moment of its actual formation, \ie
at Euclidean time  $\tau =0$. At this moment the bubble is
instantaneously at rest\cite{COL,COLDL};
the time derivative of the matter field
$\partial_\tau T\equiv
(\tau /\xi ')
\partial_{\xi'}T
$  and the metric coefficient
$\partial_\tau B\equiv
(\tau /\xi ')
 \partial_{\xi '}B\
$
both vanish at $\tau=t=0$.
As it turns out\cite{CGRII}
for   the minimal energy configuration of the bubble,
the metric coefficient $B$ and the matter field $T$ satisfy
first order differential equations. This in turn justifies
the choice that at $\tau =t=0$ the matter field $T$
and the metric coefficient $B$
of the minimal energy configuration
are only  functions of $r=\sqrt{x^2+y^2+z^2}$, \ie , the
radius of the  two-sphere.
At the moment of the actual bubble formation
one is thus working with a specific
spherically symmetric metric
Ansatz:\footnote{We calculate explicitly in the
Lorentzian instead of Euclidean signature. The
conclusion about the positive minimal energy stored
in the bubble wall or the minimal action theorem for
the bounce solution are equivalent
since time independence is assumed throughout. }
\begin{equation}
 ds^2 = B(r)(dt^2 -  dr^2 - r^2 d \Omega_2^2)
 \label{metric}
\end{equation}

For the purpose of studying  the minimal energy
configuration of the bubble wall energy density
we introduce supersymmetry charge density:
\footnote{Our conventions are the following:
 $\gamma^{\mu}=e^{\mu}_{a}\gamma^{a}$ where
$\gamma^{a}$ are the usual Dirac matrices satisfying
$\{\gamma^{a},\gamma^{b}\}=2\eta^{ab}$;
$e^{a}_{\mu}e^{\mu}_{b} = \delta^{a}_{b}$; $a=0,...3$;
$\mu=t,x,y,z$; $\overline{\psi} = \psi^{\dagger}\gamma^{t}$;
$(+,-,-,-)$ space-time signature.}
\begin{equation}
Q[\epsilon'] = 2 \int_{\partial \Sigma} d \Sigma_{\mu \nu}
(\bar \epsilon' \gamma^{\mu \nu \lambda} \psi_\lambda)
\label{susytrans}
\end{equation}
where $\Sigma$ is a space-like hypersurface enclosing the
bubble wall. Here, $\epsilon'$ is a commuting Majorana
spinor and $\psi_{\rho}$ is the spin $3/2$ gravitino field.
Taking a supersymmetry variation of $Q[\epsilon']$ with
respect to another commuting Majorana spinor
$\epsilon'$ yields
\begin{eqnarray}
\delta_{\epsilon} Q[\epsilon'] &\equiv& \{Q[\epsilon'],
\bar{Q}[\epsilon]\}
= \int_{\partial \Sigma}N^{\mu \nu} d\Sigma_{\mu \nu}\nonumber\\
&=& 2\int_{\Sigma}\nabla_{\nu}N^{\mu \nu} d\Sigma_{\mu}
\label{localchargevariation}
\end{eqnarray}
where we introduced the generalized Nester's form\cite{NESTER}
\begin{equation}
N^{\mu \nu} = \bar \epsilon'\gamma^{\mu \nu \rho}
\hat\nabla_{\rho} \epsilon\, \ . \label{nesterform}
 \end{equation}
The supercovariant derivative appearing in Nester's form is
\begin{eqnarray}
\hat\nabla_{\rho}\epsilon \equiv
\delta_{\epsilon}\psi_{\rho} &=&
[2\nabla_{\rho} + ie^{\kappa K / 2}(WP_{R} +
\bar{W}P_{L})\gamma_{\rho}\nonumber\\
 &-& Im(K_{T}\partial_{\rho}T)\gamma^{5}]\epsilon\
\label{gravitinotrans}
\end{eqnarray}
where the gravitational
derivative acting on a spinor is
$ \nabla_{\mu}\epsilon = (\partial_{\mu}
  + {1\over2}\omega^{ab}_{\mu}\sigma_{ab})\epsilon$.
In (\ref{localchargevariation})
the last equality follows
from Stoke's law.

We can describe the energy  stored in the bubble wall
(or equivalently the minimal action stored in the wall
of the bounce solution) using a thin wall
approximation\cite{COLDL} .
Such an approximation is valid in the case when the
radius $R$ of the bubble is much larger than its thickness
$2 \Delta R$, and becomes exact when $R\to \infty$.
The boundary condition on the metric coefficient is
$B(r=R)=1$, which serves as a suitable choice for
normalizing the metric. This in turn defines
the surface of the large radius bubble to be
$ 4\pi R^2$.
In the thin wall approximation, in the region
with $r\sim R$,  the metric coefficients do not change
appreciably over the range of the domain wall.
The boundary $\partial \Sigma$ are two boundaries of
two-sphere, one  at $ R -\Delta R$ and the other
at $ R+ \Delta R$, with the constraint that
$R\gg 2 \Delta R$.
In this limit, the spherical domain wall approaches
the planar domain wall.\cite{CGR}

Analysis of the surface
integral in (\ref{localchargevariation})
yields two terms: $(1)$ The ADM mass of the configuration,
denoted $ 4\pi  R^2 \cdot \sigma$
and $(2)$ The topological charge,
denoted $4\pi  R^2 \cdot \cal{C}$.
Here, $\sigma$ and $\cal{C}$ denote the ADM mass density
and the topological charge density
of the bubble wall.
The minimum topological charge density\cite{CGRII},
which corresponds
to a supersymmetric configuration,  is given by
\begin{eqnarray}
|\cal{C}|&=& \, 2 \, |\, (  \zeta
|We^{\kappa K \over 2}|)_{r=R+\Delta R}\nonumber\\
&&\quad-( \zeta
|We^{\kappa K \over 2}|)_{r=R-\Delta R}\,|
\nonumber\end{eqnarray}
\begin{equation}
={2\over {\sqrt{3\kappa}}}|
 \zeta_{R-\Delta R}
\sqrt{-V_{true}}
- \zeta_{R+\Delta R}
\sqrt{-V_{false}}|
\label{topologicalcharge}
\end{equation}
where $\zeta=\pm 1$.  $\zeta_{R-\Delta R}=-
\zeta_{R+\Delta R}$ if $W$ goes through 0 somewhere as $r$
traverses an interval $(R-\Delta R, R+\Delta R)$ and
   $\zeta_{R-\Delta R}=\zeta_{R+\Delta R}$  otherwise
(see discussion of the minimal energy solution in the
following.)
The second line in eq. (\ref{topologicalcharge}) follows from the
properties of the supersymmetric vacua, namely, as we
have shown earlier for the supersymmetric
minimum, $V=-3\kappa |W|^2e^{\kappa K}$.

The volume integral in eq. (\ref{localchargevariation})
can be shown\cite{CGR,CGRII} to be positive definite. This in turn
implies
\begin{equation}
\sigma \ge |\cal{C}|
\label{localbound}
\end{equation}
which is saturated if and only if $\delta_\epsilon
Q[\epsilon]=0$. It can be shown that
saturation of this bound implies that the bosonic
background is supersymmetric and
$\delta\psi_{\mu} = 0$ and
$\delta\chi = 0$.
For the tunnelling from  Minkowski
($V_{false}=0$) or
 AdS
($V_{false}\neq 0$)
to AdS
($V_{true}\neq 0$)
inequality (\ref{localbound}) yields:
\begin{eqnarray}
{3\over 4}\kappa
\sigma^2&\geq&
3\kappa
(|W e^{\kappa K}|_{R+\Delta R}
-|W e^{\kappa K}|_{R-\Delta R})\nonumber\\
&&\quad=(\sqrt{-V_{true}}
-\sqrt{-V_{false}})^2.\label{adsbound}
\end{eqnarray}
Inequality,  (\ref{adsbound}) , is
 the central result.
Notice that the positive energy bound
(\ref{adsbound}) for the minimal
energy density  of the thin domain wall
has precisely the opposite inequality sign  as
eq. (\ref{cdlboundII}) of the Coleman-DeLuccia
bound for the existence of a bubble instanton.
In other words, vacuum tunnelling is not allowed in
$N=1$ supergravity since the available vacuum energy
difference is not sufficient to materialize the
tunnelling bubble. For the marginal case\footnote{
In the case that the bubble forming between the two
AdS space-times has $W$ going through 0 somewhere
in between, the supersymmetric configuration satisfies
${3/ 4}\kappa
\sigma^2=
3\kappa
(|W e^{\kappa K}|_{R+\Delta R}
+|W e^{\kappa K}|_{R-\Delta R})
=(\sqrt{-V_{true}}
+\sqrt{-V_{false}})^2$
and so the Coleman-DeLuccia bound is \sl never \rm
saturated; i.e. tunnelling is super-suppressed.
Therefore, we do not discuss this case further. }
when the bound (\ref{adsbound})
is saturated, such a domain wall configuration
is supersymmetric.
However, even in this case, since the matter energy
gain by tunnelling to the true supersymmetric vacuum
is precisely cancelled by the bubble wall energy plus
the gravitational energy of the AdS space-time inside
the bubble wall, the result is an infinitely large
radius of the critical tunnelling bubble as was  shown explicitly
in \cite{CGRII}.

The  solution  in the limit $R\to\infty$
can be found by examining the equations of motion, {\it i.e.},
the first order Bogomol'nyi equations
$\delta_{\epsilon} \chi = 0$ and $\delta_{\epsilon}
 \psi_\mu = 0$ which are necessary conditions for a
supersymmetric bosonic configuration.
Technical details for the derivation
of these equations in the spherical frame
are presented in Ref.~\cite{CGRII}.
The upshot of the analysis is that the constraints on the
spinor arising from the Bogomol'nyi equations
are incompatible     unless the wall is located strictly at $R
\to\infty$. Indeed,
this corresponds to a
planar  static supersymmetric
domain wall\cite{CGR,CG},\footnote{ Examples of global supersymmetric domain
walls are given in Refs.\cite{CQR,AT}.} which can be identified
with the vacuum bubble in the
limit as the bubble radius of curvature $R\to\infty$.

In this limit  the
equations  of motion for the
matter field $T(r)$
and the metric coefficient $B(r)$
can be written as:
\begin{eqnarray}
\partial_r T(r) &=& \zeta \sqrt{B}|W|
e^{\kappa K/2}K^{T \bar T}
{D_{\overline{T}}\overline{W}\over \overline{W}},\nonumber \\
  \partial_r({1\over {\sqrt B}})&=&
\kappa\zeta |W|e^{\kappa K/ 2}.
\label{summary}
\end{eqnarray}
The constraint on the phase of $W$ along with  the
first of (\ref{summary})
imposes the geodesic equation\cite{CGR}:
\begin{equation}
Im(\partial_{r}T{D_{T}W\over W}) = 0.
\label{geodesic}
\end{equation}
This result implies that in the limit $\kappa\to 0$,
 $W(r)$ lies in the $W$ plane
 on a straight line that extends
through the origin.
Explicit solutions in $R\to\infty$ limit are identical
to those of the planar domain wall.

In the case of a planar domain wall the wall can be put
in the $(x,y)$ plane. $T(z)$ and $B(z)$ satisfy
the same eqs. (\ref{summary}), with $r$ being replaced by  $z$.
Explicit solutions for $ T(z),\ B(z)$ have been  obtained
in specific cases in Ref.~\cite{CG}.

The wall interpolating between a
 Minkowski vacuum
 ($W_{z=+\infty}=0$) and
an
AdS vacuum ($W_{z=-\infty}\neq0$) is most interesting.
These vacua have
a distinct space-time topology; Minkowski
topology is $\Re^{4}$
and AdS topology $S^{1}(time) \times \Re^{3}(space)$.
The  metric coefficient is then of the form:
\begin{eqnarray}
B(z)\to1&,&\hskip0.5cm      z\to\infty;\nonumber \\
B(z)\to {3\over{\kappa|V_{AdS}|z^2}}&,&
\hskip0.5cm z\to -\infty,
\label{madsm}
\end{eqnarray}
where $V_{AdS}=-3\kappa|W|^2e^{\kappa K}\
_{|z\to -\infty}$.

In such a domain wall background the geodesic motion of massive particles
in the z direction\footnote{
The metric is invariant under $x,y$
boosts and thus without loss we can move to an inertial frame
in which there is no motion in these directions.}
satisfies the world-line equation:\cite{CG}
\begin{equation}
({dz \over dt})^{2} + {A \over \epsilon^{2}} = 1.
\label{conservation}
\end{equation}
with $\epsilon$ being energy per mass.

A convenient way to understand massive particle motion is to
consider a particle with a given initial coordinate velocity
$v_{o}$ at some coordinate $z_{o}$;
from (\ref{conservation})
$\epsilon$ for such a particle is
$\epsilon^{2} = A(z_{o})(1-v^{2}_{o})^{-1}$.  Eq.~(\ref{conservation})
 can be thought of as the conservation of energy
with an effective potential $V(z)\equiv (1-v^2)
 = {A(z) \over A(z_{o})}(1-v_{o}^{2})$.
Again, points where $V(z) = 1$ are turning
points.

For particles incident upon the wall
 from the Minkowski
side, passage through to the AdS side is always allowed.
However, the reverse motion requires the initial velocity
to satisfy $v^{2} > 1 - A(z_{o})$; otherwise there is a turning point
and the particle returns to the AdS side.

One can understand the repulsive nature of these space-times
on the AdS side
by calculating the force on a test particle which
has a fixed position $z$ (fiducial observer).
This force can be obtained through the
geodesic equation   and yields
the magnitude of the
acceleration
$  |a|^{2} \equiv |f_{\alpha}f^{\alpha}|/m^{2}
= ({1 \over 2}{ \partial_{z}lnA \over A^{1/2} })^{2} =
(\kappa |W|e^{{\kappa K\over 2}})^{2}.$
Away from the wall,
the proper acceleration has the
constant magnitude.
In this region, integration of (\ref{conservation}) yields the
hyperbolic world line for freely falling test particles
$z^{2}-t^{2}=a^{-1}
\epsilon^{-2}$, \ie\  they are Rindler particles.
On the other hand
on the Minkowski side of the walls, free test particles
experience no gravitational force even though there is an infinite
object nearby.

One can understand the no-force result for the particles living
on the Minkowski side of the walls through the formalism of
singular hypersurfaces.\cite{ISR}  A straightforward
calculation\footnote{See \cite{HARALD} for an example of this formalism
applied to a planar geometry.}
yields a
\sl negative \rm effective gravitational mass/area  due to the wall
whereas AdS has exactly the opposite positive gravitational mass.
Thus the observer on the Minkowski side of the wall   does not
feel any gravitational force.

In conclusion  we have studied the issue of false
vacuum decay in supergravity theory.
We have found that the supersymmetric vacua are stable
against false vacuum decay into other supersymmetric vacua
nonperturbatively, \ie , to all orders in  $\kappa=8\pi
G_N$ expansion. For example, the AdS supersymmetric vacuum is
 not connected via quantum tunnelling to a
supersymmetric Minkowski vacuum.
The results are exact in $G_N$ and
 complete perturbative analysis (in the leading order of
$G_N$)  by Weinberg.\cite{WEIN}

The technique to obtain the minimal
surface energy of the
vacuum bubble and the consequent absence of tunnelling between
the supersymmetric vacua is complementary to the
positive energy theorems for supersymmetric AdS vacua.\cite{GHW}
Namely, the minimal total energy of
supersymmetric vacua is absolutely zero.\cite{GHW}
Such vacua are therefore degenerate and
consequently there can be no tunnelling. This conclusion
complements and conforms to the Bogomol'nyi bound of eq. (\ref{adsbound})
for the  minimal energy density of the bubble wall.
in this sense the problem of vacuum degeneracy in a locally

As a consequence, the degeneracy of supergravity vacua
yields static domain walls.\cite{CGR,CG}  For example, there is a
domain wall configuration interpolating between
a supersymmetric Minkowski vacuum, whose topology is
$\Re^{4}$, and a supersymmetric AdS vacuum,  whose
topology is $S^{1}$(time)$\times \Re^{3}$(space), thus
yielding a meaning to vacuum degeneracy for such topologically
distinct vacua as supersymmetric Minkowski and AdS
vacua. Study of interesting space-time effects, {\it e.g.}
 description of geodesically complete space-times and existence of Cauchy
horizons,
 in such domain wall backgrounds is
underway.\cite{CDGS}

The work presented here has been done in collaboration
with S. Griffies and S.-J. Rey. I would like to thank them
as well as R. Davis and H. Soleng for  discussions.
The    research was  supported in part by the
U.S. DOE Grant DE-AC02-76-ERO-3071,
and by a junior faculty
SSC fellowship.

\end{document}